\documentclass[twocolumn,prl,preprintnumbers,superscriptaddress,landscape,nofootinbib]{revtex4}
\usepackage[utf8]{inputenc}
\usepackage{amsmath}
\usepackage{amsthm}
\usepackage{amssymb}
\usepackage{float}
\usepackage{braket}
\usepackage{graphicx}
\usepackage{url}
\usepackage{here}
\usepackage{natbib}
\usepackage{rotating}
\usepackage{comment}
\usepackage{bm}
\usepackage[colorlinks=true, pdfstartview=FitV, linkcolor=red, citecolor=blue, urlcolor=blue]{hyperref}
\usepackage{slashed}
\usepackage{multirow}
\usepackage[normalem]{ulem}

\begin{document}

\title{Magnetic Weyl semimetals as a source 
of circularly polarized THz radiation}

\author{Jeremy Hansen}
\email[]{jchansen@iastate.edu}
\affiliation{
Department of Physics and Astronomy, Iowa State University, Ames, Iowa, 50011, USA}

\author{Kazuki Ikeda}
\email[]{kazuki.ikeda@stonybrook.edu}
\affiliation{Co-design Center for Quantum Advantage, Stony Brook University, Stony Brook, New York 11794-3800, USA}
\affiliation{Department of Physics and Astronomy, Stony Brook University, New York 11794-3800, USA}

\author{Dmitri E. Kharzeev}
\email[]{dmitri.kharzeev@stonybrook.edu.}
\affiliation{Co-design Center for Quantum Advantage, Stony Brook University, Stony Brook, New York 11794-3800, USA}

\affiliation{Department of Physics and Astronomy, Stony Brook University, New York 11794-3800, USA}

\affiliation{Department of Physics and RIKEN-BNL Research Center, Brookhaven National Laboratory,  Upton, New York 11973, USA}
\
\author{Qiang Li}
 \email[]{qiangli@bnl.gov.}
 \affiliation{Department of Physics and Astronomy, Stony Brook University, New York 11794-3800, USA}
\affiliation{Condensed Matter Physics and Materials Science Department, Brookhaven National Laboratory, Upton, NY 11973, USA}

\author{Kirill Tuchin}
\email[]{tuchink@gmail.com}
\affiliation{
Department of Physics and Astronomy, Iowa State University, Ames, Iowa, 50011, USA}

\begin{abstract}
We propose to use the electromagnetic radiation induced by a few MeV electron beam in magnetic Weyl semimetals as a source of the circularly polarized photons in the THz frequency range.
\end{abstract}

\maketitle

\emph{Introduction}. --- A fast charged particle traversing a chiral medium emits, by means of the chiral anomaly, a peculiar form of circularly polarized electromagnetic radiation known as the Chiral Cherenkov radiation \cite{Tuchin:2018sqe,Hansen:2020irw,Tuchin:2018mte}. It appears as a result of anomalous response of the chiral medium to the electromagnetic field of a fast charge. Therefore, it should appear in a variety of systems 
ranging from quark gluon plasma to Weyl and Dirac semimetals  \cite{Huang:2018hgk}. In this letter we argue that Chiral Cherenkov radiation in Weyl semimetals is a useful source of polarized electromagnetic radiation in the practically important THz frequency range.

\emph{Photon dispersion in Weyl semimetals}.---
An external electromagnetic field in Weyl semimetals induces  the anomalous Hall current
\begin{align}\label{e1}
\bm j _\text{AH}= \bm b\times \bm E\,,
\end{align}
where $\bm b=(\alpha/\pi)\bm \Delta$, with $\bm \Delta$ being the separation of Weyl nodes in momentum space and $\alpha$ the fine structure constant. A typical Weyl semimetal has more than one pair of nodes in which case $\bm \Delta$ stands for an effective displacement. It was suggested in \cite{Zubkov:2023dpe} that the total displacement is the vector sum of all displacements: $\bm \Delta=\sum_i \bm \Delta_i$. In this case $\bm \Delta$ is finite only in magnetic Weyl semimetals but vanishes in non-magnetic ones. For example, in Co$_3$Sn$_2$S$_2$ $b=3.8$~eV \cite{Zubkov:2023dpe,wang2018large}.

\begin{figure}
    \centering
    \includegraphics[width=\linewidth]{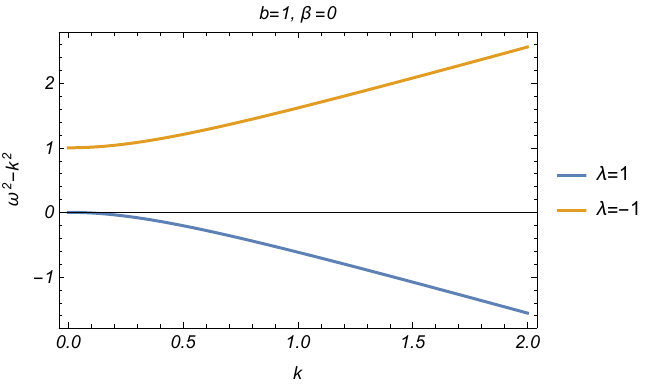}
    \caption{Chiral branches of the photon dispersion relation. Arbitrary units.}
    \label{fig:branches}
\end{figure}

\begin{figure}[H]
    \centering
    \includegraphics[width=0.5\linewidth]{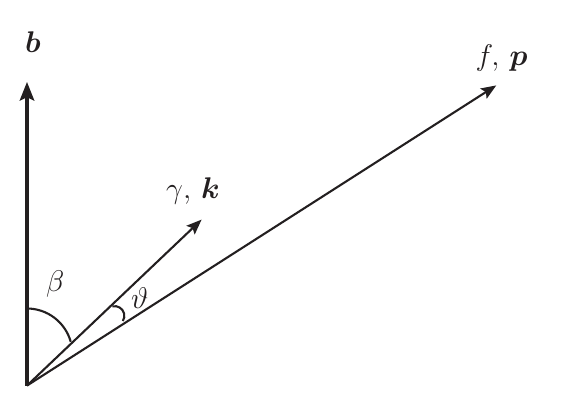}
    \caption{The geometry of the photon radiation. The three vectors are generally not in the same plane, however the azimuthal angle is not shown. At high energy $\vartheta\ll \beta$, i.e.\ $\beta$ is approximately the angle between $\bm p$ and $\bm b$ (angle of incidence).  }
    \label{fig:geometry}
\end{figure}

In the presence of the current (\ref{e1}) the dielectric tensor takes form $\varepsilon_{ij}= \varepsilon\delta_{ij}- i\epsilon_{ijk}b_k/\omega$, where $\varepsilon \approx 1/(1+\omega_p^2/\omega^2)$ for $\omega\gg \omega_p$. The photon dispersion relation is a solution of the Fresnel equation
\begin{align}\label{f1}
\left|k_ik_j-k^2\delta_{ij}+\omega^2\varepsilon_{ij}\right|=0\,
\end{align}
in units where $\hbar =c=1$, which we assume henceforth. The dispersion relation has the general form $\omega^2-k^2=\mu^2(\bm k,\lambda)$ where $\bm k$ is the wave vector and, setting $\omega_p=0$ for the sake of presentation simplicity, \cite{Qiu:2016hzd}  
\begin{align}\label{e3}
\mu^2(\bm k,\lambda)=\frac{b^2}{2}-\lambda\, \text{sgn}(\bm k\cdot \bm b)
\sqrt{(\bm k\cdot \bm b)^2+b^2/4}\,.
\end{align}
Here $\lambda=\pm 1$ is the photon circular polarization. Let $\beta$ be the angle between $\bm b$ and $\bm k$, see Fig.~\ref{fig:geometry}.
There are two branches of the dispersion relation: (i) the gapped branch:  $\lambda\cos\beta>0$, $\mu^2>0$  and (ii) the gapless branch $\lambda\cos\beta<0$, $\mu^2<0$. These are depicted in Fig.~\ref{fig:branches}. Photon radiation is associated with the latter. 

\emph{Photon radiation}.--- 
The initial and final states of electron are solutions to the Dirac equation with energies $E$, $E'$ and  momenta $\bm p$, $\bm p'$. The  wave function of the radiated photon with energy and momentum $\omega$, $\bm k$ and polarization $\lambda$ reads \cite{alekseev1966quantization}
\begin{align}\label{g1}
\bm A= &\bm e \left( \frac{k\, v}{2\omega^2\varepsilon_{ij}e^*_ie_{j}}\right)^{1/2} e^{i\bm k \cdot \bm x-i\omega t}
\end{align}
where $\bm e$ is the polarization vector for a given $\lambda$ and $v=\partial \omega/\partial k$ is the group velocity. The explicit expression for $\bm e$ can be found in \cite{Tuchin:2018mte}. Computing the transition amplitude $e^-\to e^-\gamma$ and using the standard procedure to obtain the transition rate we obtain for the number of photons emitted per unit time in unit solid angle in the  frequency interval $d\omega$ \cite{Tuchin:2018mte}: 
\begin{align}\label{g2}
\frac{dW}{d\Omega d\omega}=& \frac{\alpha}{16\pi}\sum_\lambda\delta(\omega+E'-E)\frac{k^3}{EE'\omega^2\varepsilon_{ij}e_{i}^*e_{ j}}\nonumber\\
&4e^*_{i}e_{ j}\left[ p_ip_j'+p_jp_i'+\delta_{ij}(EE'-\bm p\cdot \bm p'-m^2)\right]\,,
\end{align}
where $\bm p'= \bm p -\bm k$. The energy conservation expressed by the delta-function is satisfied only for the gapless branch $\mu^2<0$. Similarly to the conventional Cherenkov radiation, the Chiral Cherenkov one is radiated even by a particle moving with constant velocity, which corresponds to the limit $\omega\ll E$ when electron's recoil is small.  In this case, the radiating electron moves in phase with the wave. As the result both types of Cherenkov radiation have infinite coherence length in contrast to the bremsstrahlung. 

Thus far we discussed the Chiral Cherenkov radiation by a single electron. 
A typical electric current pulse consists of $N=10^{11}$ electrons (equivalent to the electron bunch charge 16 nC). The corresponding power $dI=N dP$.

The explicit form of the rate is rather bulky. We therefore consider two special cases.

\begin{figure}
    \centering
    \includegraphics[width=\linewidth]{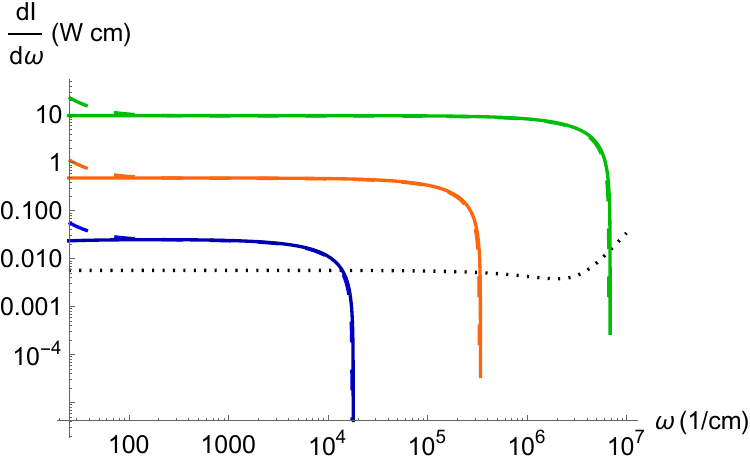}
    \caption{Intensity spectrum of Chiral Cherenkov radiation by a pulse of $N=10^{11}$ electrons of energy $E=3$~MeV, $\beta=0$ in a Weyl semimetal with $ b=3.8$~eV (green), $b=0.19$~eV (orange), and $ b=0.01$~eV (blue). Solid line: Eq.~(\ref{e6}), dashed line: Eq.~(\ref{g20}) with $\omega_p=0.6$~meV, dotted line: bremsstrahlung spectrum in Co$_3$Sn$_2$S$_2$ computed according to \cite{Workman:2022ynf}. 1/cm $\approx$ 0.03~THz. }
    \label{fig:spectra1}
\end{figure}

\begin{figure}
    \centering
    \includegraphics[width=\linewidth]{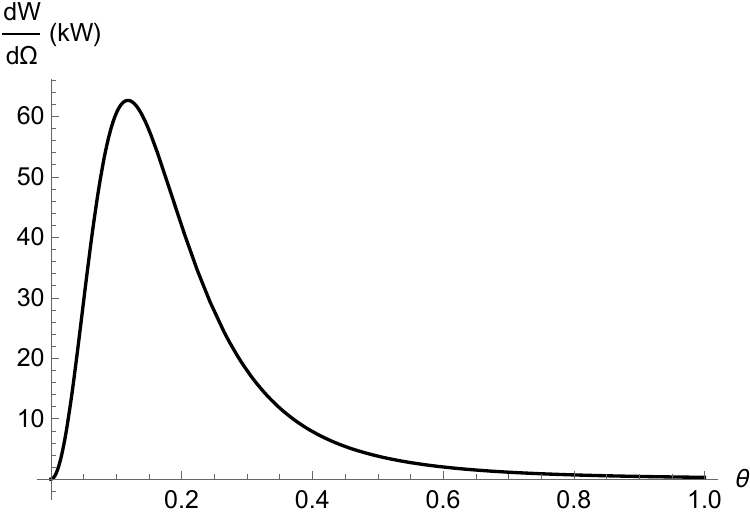}
    \caption{Angular distribution of  Chiral Cherenkov radiation.  $N=10^{11}$, $E=3$~MeV, $\beta=0$, $b=3.8$~eV. Eq.~(\ref{e6}) and Eq.~(\ref{g20}) are indistinguishable.}
    \label{fig:spectra2}
\end{figure}

\begin{figure}
    \centering
    \includegraphics[width=\linewidth]{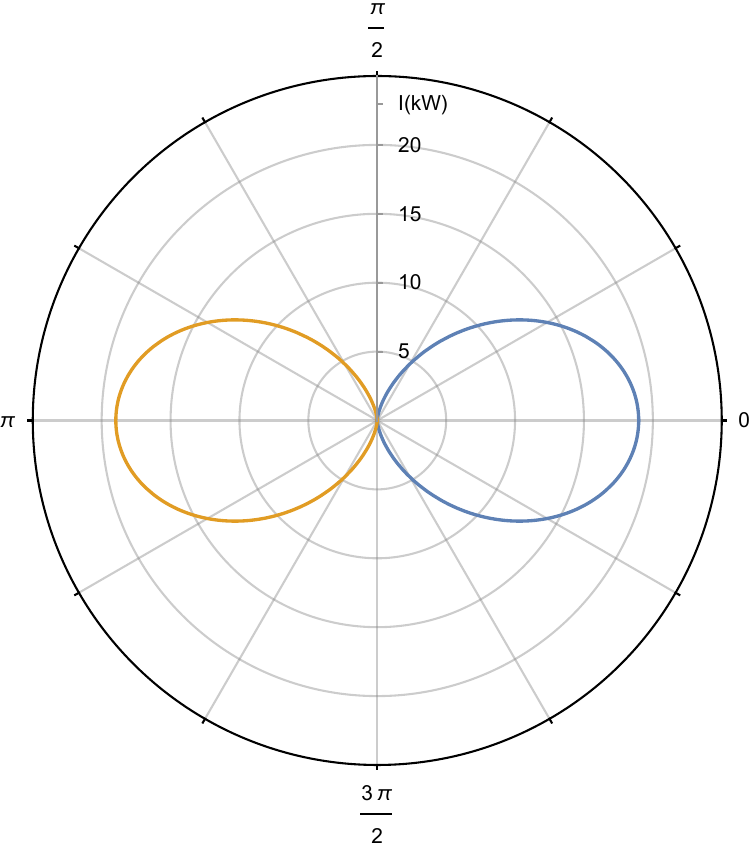}
    \caption{Variation of Chiral Cherenkov radiation intensity with electron angle of incidence (the polar angle). The two loops have opposite polarization.
   $N=10^{11}$, $E=3$~MeV, $b=3.8$~eV. }
    \label{fig:spectra3}
\end{figure}

\emph{High energy approximation}.--- 
Assume that $E\gg m$ and $\omega\gg |\mu|$. The latter condition is equivalent to $\omega\gg b,\omega_p$. The Chiral Cherenkov radiation rate (\ref{g2}) can then be written in a compact form
\cite{Tuchin:2018sqe,Tuchin:2018mte,Hansen:2020irw}:
\begin{align}\label{e6}
\frac{dW}{d\omega d\Omega}=& \frac{\alpha\omega}{2\pi E}\delta(\omega^2\vartheta^2+\kappa)\nonumber\\
&\times\left\{ f E\left( \frac{\omega^2}{2E^2}-\frac{\omega}{E}+1\right)-m^2 \frac{\omega}{E}\right\}\theta(-\kappa)\,,
\end{align}
where $f=\lambda b \cos\beta$, and $\beta$ is now  the angle between the incident electron direction and the vector $\bm b$  since in this approximation the angle $\vartheta$ between $\bm k$ and $\bm p$ is small, see Fig.~\ref{fig:geometry}. 
In the right-hand-side $\theta$ is the step function and we defined
\begin{align}\label{e8}
\kappa= -\left(1-\frac{\omega}{E}\right)\omega f+\frac{m^2\omega^2}{E^2}\,. 
\end{align}
The kinematic condition $\kappa<0$ can be satisfied only if $f>0$. Since  $\mu^2\approx -f\omega$ it is equivalent  to $\mu^2 <0$. Additionally, it imposes the upper limit on the photon spectrum:
\begin{align}\label{e9}
\omega<E(1+m^2/fE)^{-1}\,.
\end{align}
 The lower limit is proportional to $\omega_p$, but in the high-energy approximation it is neglected. The differential spectrum of the radiation power is given by 
\begin{align} \label{e10}
\frac{dP}{d\omega d\Omega} = \omega
\frac{dW}{d\omega d\Omega}
\end{align} 
solid lines in Figs.~\ref{fig:spectra1},\ref{fig:spectra2},\ref{fig:spectra3} display the high-energy approximation of the radiated photon spectrum, its angular distribution and dependence on angle of incidence (viz.\ angle between the electron momentum $\bm p$ and the vector $\bm b$). 

The photon emission rate Eq.~(\ref{e6}) can be further simplified by noting that in semimetals $m^2\gg fE$:
\begin{align}\label{e11}
\frac{dW}{d\omega d\Omega}\approx \frac{\alpha f}{2\pi \omega}\delta\left(\vartheta^2-\frac{f}{\omega}+\frac{m^2}{E^2}\right)
\theta\left(\frac{E^2 f}{m^2}-\omega\right)\,,
\end{align}
where the step-function follows from (\ref{e9}). Multiplying by $\omega$ and integrating over $d\Omega = \pi d\vartheta^2$ we obtain the power  spectrum 
\begin{align}\label{g11}
\frac{dP}{d\omega} = \frac{\alpha f}{2}\theta\left(\frac{E^2 f}{m^2}-\omega\right)\,,
\end{align}
and the the total power 
\begin{align}\label{g12}
P=\frac{\alpha f^2E^2}{2m^2}= \frac{\alpha b^2\cos^2\beta E^2}{2m^2}.
\end{align}
Integrating (\ref{e11}) over $\omega$ furnishes the angular distribution of the radiation
\begin{align}\label{g13}
\frac{dW}{d\Omega}=\frac{\alpha f}{2\pi}\frac{1}{\vartheta^2+m^2/E^2}\,.
\end{align}

Evidently, the total radiation power increases as $b^2$, while the UV cutoff of the spectrum as   $b$. It is therefore advantageous to have as high $b$ as possible to obtain the highest power of the Chiral Cherenkov radiation. It also guarantees, as per Figs.~\ref{fig:spectra1},\ref{fig:spectra2}, that the radiation power of the polarized photons by means of the Chiral Cherenkov mechanism is larger than that of the non-polarized photons through the conventional sources.

\emph{Electron incident along $\bm b$}.--- 
According to (\ref{g12}) the radiation power vanishes when the incident electron moves perpendicularly to the vector $\bm b$. Therefore, in practice one would like to position the Weyl semimetal sample such that $\bm p\times \bm b=0$. In this case (\ref{g2}) significantly simplifies. Perfoming the angular integral yields the rate:
\begin{align}\label{g20}
    \frac{dW}{d\omega}=&\frac{(\omega^2-\mu^2)(E E'-m^2)-(2\omega E-\mu^2)(2\omega E'+\mu^2)}{4(\omega^2-\mu^2)\omega^2E |\bm p|}\nonumber\\
    &\times\frac{\alpha}{2}\left(1+\omega_p^2/\omega^2\right)\theta(\omega_+-\omega),
\end{align}
where $\mu^2=\omega_p^2-\lambda b \omega$, $\bm p^2=E^2-m^2$ and 
\begin{align}
\omega_+=&\bigg[\lambda  b (2 \bm p^2+  \omega_p^2)+2 E \omega_p^2+ \nonumber\\
&
 +2|\bm p|\sqrt{\left( \lambda b E -\omega_p^2\right)^2-m^2 \left(b^2 +4 \omega_p^2\right)}\bigg]\nonumber\\
\times&\left[ \lambda b ( 4E+\lambda b )+4 m^2\right]^{-1}.
\end{align}
In the high energy limit $\omega\gg b$ and  $E\gg m$ these equations reduce to those displayed in the previous section.  
Dashed lines in Figs.~\ref{fig:spectra1},\ref{fig:spectra2} represent a calculation of intensity with Eq.~(\ref{g20}). Evidently, the high-energy approximation is quite accurate all the way down to the IR region $\omega\sim \omega_p$ where our approximation for the dielectric constant $\varepsilon$ breaks down anyway.

%

\emph{Dirac semimetals }.---In Dirac semimetals $\bm b=0$. However, by applying external parallel magnetic $B$ and electric $\mathcal{E}$ fields one can generate the chiral charge. The corresponding axial chemical potential reads \cite{Fukushima:2008xe}: 
\begin{equation}
\label{eq:new_mu5}
    \mu_5=\frac{3}{4}\frac{v^3}{\pi^2}e^2\frac{\mathcal{E}B}{T^2}\tau_V,
\end{equation} 
where $\tau_V$ is the chirality relaxation time, and $v$ is the Fermi velocity of Dirac quasiparticles.
The chiral anomaly manifests itself as the Chiral Magnetic Effect which is induction of electric current in the direction of the applied magnetic field:
\begin{align}\label{f2}
\bm j_\text{CM}= \sigma_\chi \bm B\,,
\end{align}
where the chiral magnetic conductivity is $\sigma_\chi = (2/\pi)\alpha \mu_5$.

In the presence of the current (\ref{f2}) the photon dispersion relation acquires a branch with $\mu^2<0$ in a way very similar to Weyl semimetals. In fact all formulas derived in the high-energy limit, apply at finite $\mu_5$ as well upon the substitution  $f=\lambda\sigma_\chi$ \cite{Tuchin:2018sqe,Hansen:2020irw}.

\begin{figure}
    \centering
    \includegraphics[width=\linewidth]{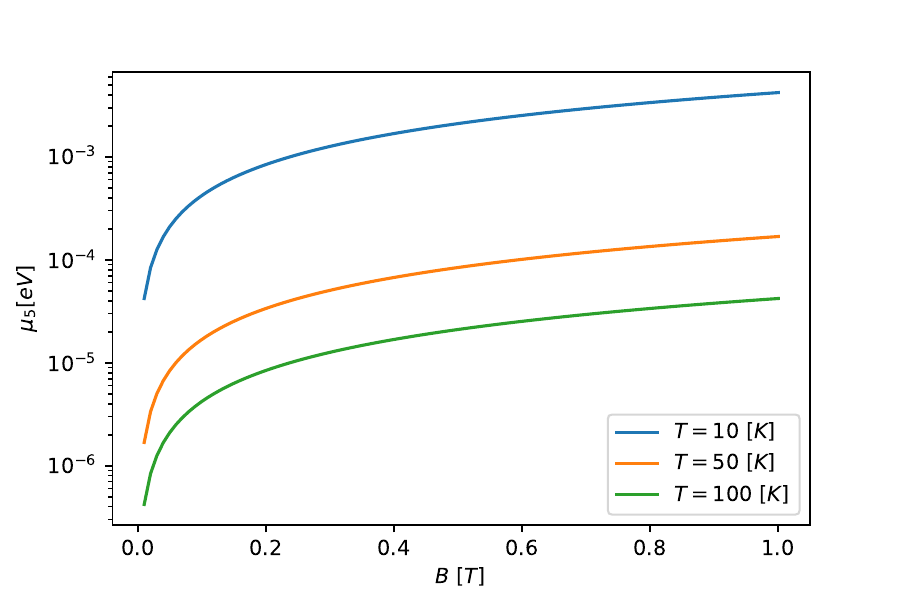}
    \caption{The temperature and magnetic field dependence of $\mu_5$ at $\mathcal{E}=1~mV/m$. The chiral magnetic conductivity is $\sigma_\chi = 4.6\cdot 10^{-3}\mu_5$. }
    \label{fig:enter-label}
\end{figure}

In practice, however, the values of $\sigma_\chi$ are much smaller than $b$ as can be seen in Fig.~\ref{fig:enter-label}.
The separation of the Weyl nodes in a chiral semimetal is of the order $\Delta  \sim 10^{-2} (\pi/a)$. For example, in  TaAs it is $\Delta =80$~eV \cite{Xu:2015cga,Lv:2015pya}. The corresponding $b$-parameter in (\ref{e1}) is of magnitude $b=0.19$~eV. In contrast, in a magnetic Weyl semimetal Co$_3$Sn$_2$S$_2$ $b=3.8$~eV \cite{Zubkov:2023dpe,wang2018large}.
\vskip0.3cm

\emph{Summary.}--- 
     The Chiral Cherenkov radiation appears due to anomalous response of chiral media to the electromagnetic field of a fast charge.
    In magnetic Weyl semimetals, for an intense electron beam of a few MeV  energy, the intensity of the Chiral Chernkov radiation can exceed the intensity of usual bremsstrahlung in the THz frequency range.
    Moreover, the THz CC radiation is circularly polarized, which can be very useful for practical applications.

\section*{Acknowledgements}
We thank Sebastian Grieninger, Fangcheng He and Andrea Palermo for a useful discussion. This work was supported by the U.S. Department of Energy, Office of Science, National Quantum Information Science Research Centers, Co-design Center for Quantum Advantage (C2QA) under Contract No.DE-SC0012704 (KI, DK), and the U.S. Department of Energy under Grants DE-FG88ER41450 (DK) and DE-SC0012704 (DK), This work of JH and KT was supported in part by the U.S. Department of Energy Grant No.\ DE-SC0023692. The work of QL was supported by the U.S. Department of Energy, Office of Basic Energy Sciences, Division of Materials Sciences
and Engineering, under contract No. DE-SC0012704.

\bibliographystyle{utphys}
\bibliography{ref}

\providecommand{\href}[2]{#2}\begingroup\raggedright\begin{thebibliography}{10}

\bibitem{Tuchin:2018sqe}
K.~Tuchin, ``{Radiative instability of quantum electrodynamics in chiral
  matter},'' \href{http://dx.doi.org/10.1016/j.physletb.2018.09.055}{{\em Phys.
  Lett. B} {\bfseries 786} (2018) 249--254},
  \href{http://arxiv.org/abs/1806.07340}{{\ttfamily arXiv:1806.07340
  [hep-ph]}}.

\bibitem{Hansen:2020irw}
J.~Hansen and K.~Tuchin, ``{Collisional energy loss and the chiral magnetic
  effect},'' \href{http://dx.doi.org/10.1103/PhysRevC.104.034903}{{\em Phys.
  Rev. C} {\bfseries 104} no.~3, (2021) 034903},
  \href{http://arxiv.org/abs/2012.06089}{{\ttfamily arXiv:2012.06089
  [hep-ph]}}.

\bibitem{Tuchin:2018mte}
K.~Tuchin, ``{Chiral Cherenkov and chiral transition radiation in anisotropic
  matter},'' \href{http://dx.doi.org/10.1103/PhysRevD.98.114026}{{\em Phys.
  Rev. D} {\bfseries 98} no.~11, (2018) 114026},
  \href{http://arxiv.org/abs/1809.08181}{{\ttfamily arXiv:1809.08181
  [hep-ph]}}.

\bibitem{Huang:2018hgk}
X.-G. Huang and K.~Tuchin, ``{Transition Radiation as a Probe of the Chiral
  Anomaly},'' \href{http://dx.doi.org/10.1103/PhysRevLett.121.182301}{{\em
  Phys. Rev. Lett.} {\bfseries 121} no.~18, (2018) 182301},
  \href{http://arxiv.org/abs/1808.00635}{{\ttfamily arXiv:1808.00635
  [hep-ph]}}.

\bibitem{Zubkov:2023dpe}
M.~A. Zubkov, ``{Weyl orbits as probe of chiral separation effect in magnetic
  Weyl semimetals},'' \href{http://arxiv.org/abs/2311.12712}{{\ttfamily
  arXiv:2311.12712 [cond-mat.mes-hall]}}.

\bibitem{wang2018large}
Q.~Wang, Y.~Xu, R.~Lou, Z.~Liu, M.~Li, Y.~Huang, D.~Shen, H.~Weng, S.~Wang, and
  H.~Lei, ``Large intrinsic anomalous hall effect in half-metallic ferromagnet
  co3sn2s2 with magnetic weyl fermions,'' {\em Nature communications}
  {\bfseries 9} no.~1, (2018) 1--8.

\bibitem{Qiu:2016hzd}
Z.~Qiu, G.~Cao, and X.-G. Huang, ``{On electrodynamics of chiral matter},''
  \href{http://dx.doi.org/10.1103/PhysRevD.95.036002}{{\em Phys. Rev. D}
  {\bfseries 95} no.~3, (2017) 036002},
  \href{http://arxiv.org/abs/1612.06364}{{\ttfamily arXiv:1612.06364
  [cond-mat.mes-hall]}}.

\bibitem{alekseev1966quantization}
A.~Alekseev and Y.~P. Nikitin, ``Quantization of the electromagnetic field in a
  dispersive medium,'' {\em Sov. Phys. JETP} {\bfseries 23} (1966) 608.

\bibitem{Workman:2022ynf}
{\bfseries Particle Data Group} Collaboration, R.~L. Workman and Others,
  ``{Review of Particle Physics},''
  \href{http://dx.doi.org/10.1093/ptep/ptac097}{{\em PTEP} {\bfseries 2022}
  (2022) 083C01}.

\bibitem{Fukushima:2008xe}
K.~Fukushima, D.~E. Kharzeev, and H.~J. Warringa, ``{The Chiral Magnetic
  Effect},'' \href{http://dx.doi.org/10.1103/PhysRevD.78.074033}{{\em Phys.
  Rev. D} {\bfseries 78} (2008) 074033},
  \href{http://arxiv.org/abs/0808.3382}{{\ttfamily arXiv:0808.3382 [hep-ph]}}.

\bibitem{Xu:2015cga}
S.~Y. Xu {\em et~al.}, ``{Discovery of a Weyl Fermion semimetal and topological
  Fermi arcs},'' \href{http://dx.doi.org/10.1126/science.aaa9297}{{\em Science}
  {\bfseries 349} (2015) 613--617},
  \href{http://arxiv.org/abs/1502.03807}{{\ttfamily arXiv:1502.03807
  [cond-mat.mes-hall]}}.

\bibitem{Lv:2015pya}
B.~Q. Lv {\em et~al.}, ``{Experimental discovery of Weyl semimetal TaAs},''
  \href{http://dx.doi.org/10.1103/PhysRevX.5.031013}{{\em Phys. Rev. X}
  {\bfseries 5} no.~3, (2015) 031013},
  \href{http://arxiv.org/abs/1502.04684}{{\ttfamily arXiv:1502.04684
  [cond-mat.mtrl-sci]}}.

\end{thebibliography}\endgroup

\end{document}